\documentclass{osa-article}

\journal{osajournal}

\articletype{Letter}
\usepackage{amsmath}
\usepackage{physics}

\begin{document}

\title{Ultrafast 3.5\,\textmu m fiber laser}

\author{Nathaniel Bawden,\authormark{1,*} Ori Henderson-Sapir,\authormark{1,2} Stuart D. Jackson,\authormark{3} and David J. Ottaway\authormark{1}}

\address{\authormark{1}Department of Physics and Institute of Photonics and Advanced Sensing, The University of Adelaide, Adelaide, SA, Australia\\
\authormark{2}Mirage Photonics, Oaklands Park, SA, Australia\\
\authormark{3}MQ Photonics, School of Engineering, Faculty of Science and Engineering, Macquarie University, North Ryde, NSW, Australia}

\email{\authormark{*}nathaniel.bawden@adelaide.edu.au} 



\begin{abstract}
We report the first mode-locked fiber laser to operate in the femtosecond regime well beyond 3\,\textmu m. The laser uses dual-wavelength pumping and non-linear polarisation rotation to produce 3.5\,\textmu m wavelength pulses with minimum duration of 580\,fs at a repetition rate of 68\,MHz. The pulse energy is 3.2\,nJ, corresponding to a peak power of 5.5\,kW. 
\end{abstract}


\section{Introduction}


Lasers operating around 3.5\,\textmu m are of great interest because they directly excite and probe C-H and N-O vibrational bonds. This makes them an interesting source for polymer processing \cite{frayssinousResonantPolymerAblation2018}, molecular spectroscopy \cite{adlerMidinfraredFourierTransform2010} and breath analysis \cite{wangBreathAnalysisUsing2009}. Prior to this work, sub-picosecond pulses at this wavelength have only been achieved by using non-linear techniques to frequency shift shorter wavelength lasers \cite{schliesserMidinfraredFrequencyCombs2012a, duvalWattlevelFiberbasedFemtosecond2016}. Fiber based sources offer several potential advantages over conventional methods of producing ultrafast pulses in the mid-infrared wavelength region using OPOs and OPAs \cite{jacksonFiberbasedSourcesCoherent2020}. These include greater average output power, improved beam quality, robustness and affordability. Ultrafast fiber systems could therefore see broader implementation and use in potential new applications.

Previous demonstrations of mode-locked fiber lasers at 3.5\,\textmu m have produced picosecond scale pulses. The first demonstration by Qin et al.\ used a black phosphorous saturable absorber to mode-lock with a pulse energy of 1.38\,nJ at 28.91\,MHz \cite{qinBlackPhosphorusQswitched2018}. A pulse duration of 3.8\,ps was estimated from spectral broadening assuming a transform limited temporal Gaussian shape. We have previously used the frequency shifted feedback technique to achieve tunable pulsed operation, producing pulses as short as 53\,ps with energies up to 1.38\,nJ over a 215\,nm wavelength range \cite{henderson-sapirModelockedTunableFiber2020}.



At the time of writing, the majority of ultrafast mid-infrared fibre laser research has focused on the 3\,\textmu m band. Passive mode-locking has been achieved using several different saturable absorber materials, producing picosecond scale pulses with peak powers around the 1\,kW level \cite{zhuPulsedFluorideFiber2017}. The greatest peak power of 1.86\,kW (25\,ps pulse duration) was demonstrated by Tang et al.\ using a semiconductor saturable absorber mirror \cite{tangWattlevelPassivelyModelocked2015}. While these systems have the advantage of a simpler resonator configuration, shorter pulses with higher peak power have been produced using the more complex non-linear polarisation rotation (NPR) technique \cite{jacksonFiberbasedSourcesCoherent2020,zhuPulsedFluorideFiber2017}, which utilizes polarisation controlling optics to create an intensity dependent loss resulting in a preference for a short pulse rather then CW operation \cite{hausModelockingLasers2000}. With the inclusion of additional amplifier and non-linear stages, NPR mode-locked fiber lasers have shown themselves to be strong candidates for generating mid-infrared frequency combs and supercontinuum sources \cite{huangSubtwocycleOctavespanningMidinfrared2020, hudsonAllfiberSupercontinuumSpanning2017, woodwardGeneration70fsPulses2017, duvalWattlevelFiberbasedFemtosecond2016}.

Fiber laser development generally follows a logical progression for any particular lasing transition. Continuous wave demonstrations are followed by Q-switching and mode-locking using saturable absorbers. Ultrashort pulse generation is then achieved using NPR. The first systems to reach this point in the 3\,\textmu m wavelength range utilised the 2.8\,\textmu m transition in erbium \cite{huUltrafastPulsesMidinfrared2015, duvalFemtosecondFiberLasers2015}. Holmium-praseodymium co-doped systems followed shortly after, with the 2.9\,\textmu m operating wavelength avoiding the strong water vapor present around 2.8\,\textmu m \cite{antipovHighpowerMidinfraredFemtosecond2016}. Ultrashort pulses have recently been demonstrated using dysprosium doped fiber \cite{wangUltrafastDy3Fluoride2019}. Wang et al.\ reported pulses as short as 828\,fs with a peak power of 4.2\,kW operating near 3.1\,\textmu m, the longest wavelength sub-picosecond fiber laser prior to this work.

In this work, we advance the central operating wavelength of femtosecond fiber lasers to 3.5\,\textmu m, well within the 3\,\textmu m to 5\,\textmu m atmospheric window. The NPR mode-locked system uses an erbium-doped zirconium fluoride fiber and the dual-wavelength pumping method originally developed by Henderson-Sapir et al.\ \cite{henderson-sapirMidinfraredFiberLasers2014}. Direct operation at this wavelength range gives a high brightness, mode-locked spectrum which could be used to directly probe the absorption lines of specific molecules such as $\text{CH}_4$ and HCl, opening up possible applications in precision spectroscopy and remote sensing \cite{schliesserMidinfraredFrequencyCombs2012a, jacksonHighpowerMidinfraredEmission2012}.

\section{Experimental setup and results}
\begin{figure}[htbp]
    \centering
    \includegraphics[width=7cm]{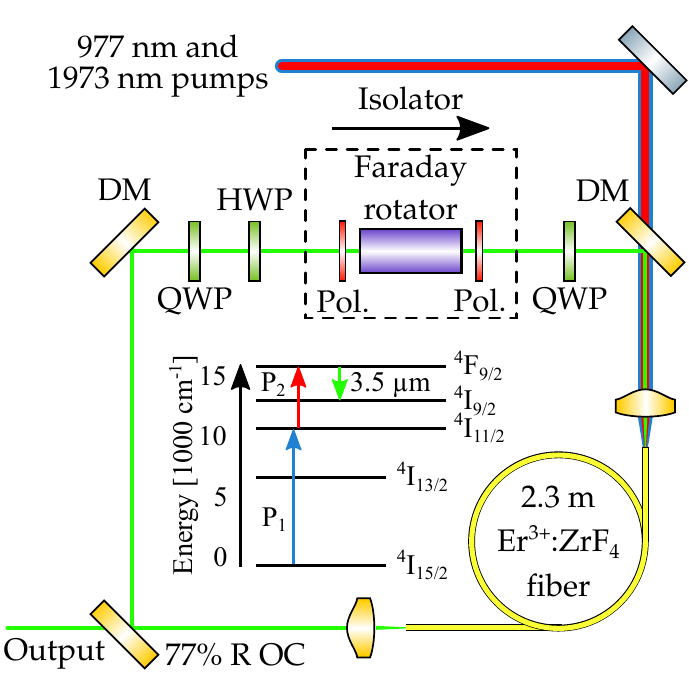}
    \caption{Schematic diagram of the NPR mode-locked laser. Inset: A simplified energy level diagram showing the dual-wavelength pumping scheme. HWP - half wave plate. QWP - quarter wave plate. DM - dichroic miror. OC - output coupler. $\text{P}_1$ - 977\,nm pump. $\text{P}_2$ - 1973\,nm pump.}
    \label{fig:setup}
\end{figure}

The experimental setup for NPR mode-locking experiments at 3.5\,\textmu m is shown in Fig.\ \ref{fig:setup}. This configuration is based on those used in the first NPR mode-locking experiments at 3\,\textmu m and adapted for 3.5\,\textmu m \cite{huUltrafastPulsesMidinfrared2015, duvalFemtosecondFiberLasers2015}.
The 977\,nm and 1973\,nm pump sources are a commercial fiber-coupled laser diode (LIMO HLU30F200-977) and a thulium-doped silica fiber laser (Mirage Photonics TFL-2000). The combined pumps are passed through a 45 degree dichroic mirror and launched into the 1\,mol\% $\text{Er}^{3+}\text{:ZrF}_4$ fiber by a custom aspheric lens. The 2.3\,m long, double-clad fiber (Le Verre Fluor\'{e}) has a 16\,\textmu m diameter core with a 0.12 NA at 3.5\,\textmu m and a 240/260\,\textmu m double-truncated circular inner cladding. The 977\,nm and 1973\,nm pumps are launched into the inner cladding and core, respectively. The output end of the fiber is angle polished to prevent back reflections and parasitic lasing at 2.8\,\textmu m. Light from this end is collimated by another aspheric lens and incident on a 45 degree output coupler mirror with 77\% reflectivity at 3.5\,\textmu m. A second 45 degree dichroic mirror then completes the ring laser resonator. The length of the free space section in the laser is approximately 0.9\,m long.

A free space Faraday rotator (Thorlabs) and two high power polarisers (FastPulse Technology) form an optical isolator that is placed between the two dichroic mirrors, providing unidirectional operation. The transmission of the complete isolator system is measured to be 86\% at 3.5\,\textmu m for linearly polarised light. Two quarter wave plates and a half wave plate (Thorlabs) are also added as shown in Fig.\ \ref{fig:setup} to implement NPR mode-locking. The quarter wave plate at the input end of the isolator is present to reduce the loss at the following polariser and was necessary to achieve mode-locked operation. The laser output is passed through a 3\,\textmu m longpass filter to remove the residual pump light when making measurement of the mode-locked system.

For a 977\,nm pump power of 6.8\,W and a 1973\,nm pump power of 5.5\,W, mode-locked operation can be started with careful adjustment of the wave plates and knocking one of the resonator mirrors which creates an intensity spike that can preferentially pass through the polarisation controlling optics and the fiber. The average output power is measured to be 216\,mW during mode-locked operation at these pump powers. A 2.4\,\textmu m to 3.2\,\textmu m bandpass filter and grating spectrometer were used to check for parasitic lasing at 2.8\,\textmu m. No lasing at this wavelength was observed during mode-locked operation. 

\begin{figure}[htbp]
         \centering
         \includegraphics[width=7cm]{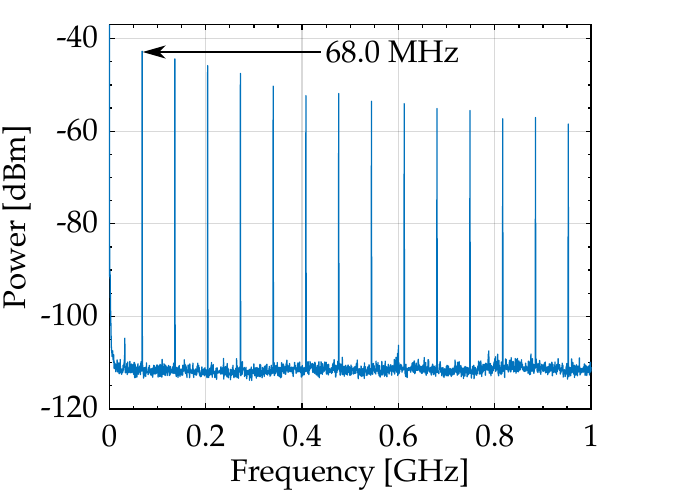}
         \caption{RF spectrum of the mode-locked laser output with a fundamental repetition rate of 68.0\,MHz.}
         \label{fig:frequency spectrum}
\end{figure}

Time domain measurements of the mode-locked output made with a 1\,ns rise time photodetector (VIGO PEM-10.6) show continuous pulsed operation. 
A peak at 68.0\,MHz is displayed in the RF spectrum, equal to the inverse round trip time of the ring resonator, with harmonics extending up to the 1\,GHz limit of the spectrum analyser (Fig.\ \ref{fig:frequency spectrum}). The fundamental beat frequency has a signal to noise ratio of nearly 70\,dB for a RBW of 3\,kHz, comparable to NPR mode-locked systems around 3\,\textmu m \cite{huUltrafastPulsesMidinfrared2015, antipovHighpowerMidinfraredFemtosecond2016}. An average pulse energy of 3.2\,nJ is calculated from the output power and repetition rate.

\begin{figure}[htbp]
    \centering
    \includegraphics[width=7cm]{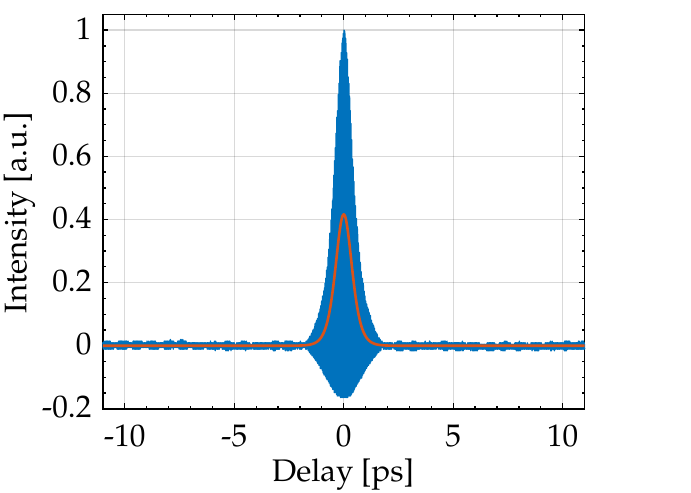}
    \caption{Autocorrelation trace of the mode-locked pulses at an average output power of 216\,mW. The corresponding pulse width is 580\,fs FWHM for a $\text{sech}^2$ shaped pulses. The 977\,nm and 1973\,nm pump powers were 6.8\,W and 5.5\,W, respectively.}
    \label{fig:autocorrerlation}
\end{figure}

An autocorrelator based on two-photon absorption was constructed to measure the pulse width (same as in \cite{henderson-sapirModelockedTunableFiber2020}). The fit to the autocorrelation trace in Fig.\ \ref{fig:autocorrerlation} gives a FWHM of 580\,fs for a $\text{sech}^2$ shaped pulse after deconvolution. A peak power of 5.5\,kW is obtained based on these values, on par with the results obtained at 2.8\,\textmu m and 3.1\,\textmu m \cite{duvalFemtosecondFiberLasers2015,wangUltrafastDy3Fluoride2019}.

\begin{figure}[htbp]
    \centering
    \includegraphics[width=7cm]{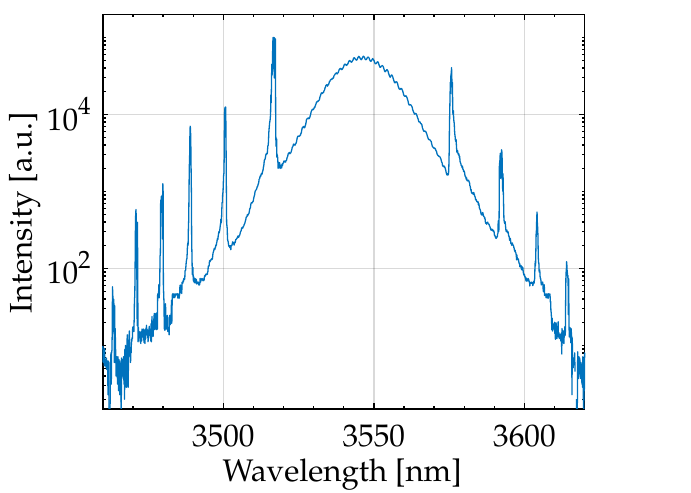}
    \caption{Optical spectrum of the mode-locked laser at maximum pump power showing Kelly sidebands. The spectrum broadens to 23\,nm FWHM.}
    \label{fig:wavelength specrtum}
\end{figure}

The optical spectrum of the mode-locked laser was measured using a grating spectrometer (Fig.\ \ref{fig:wavelength specrtum}). The operating wavelength is centred on 3.54\,\textmu m and broadens to 23\,nm FWHM when mode-locking compared to the 0.5\,nm linewidth typically measured during CW operation. The time-bandwidth product for these pulses is 0.32 indicating that they are close to transform limited (0.315 for $\text{sech}^2$).

The narrow peaks in the optical spectrum either side of the central wavelength indicate that this system operates in the soliton regime. The spacing of the sidebands from the center frequency can be used to calculate the resonator dispersion using the equation \cite{smithSidebandGenerationPerturbations1992}:
\begin{equation}
    (f_m-f_c)^2 = \frac{m}{\pi\abs{B_2}}-\frac{1}{4\pi^2\tau^2}
\end{equation}
where $m$ is the order of the sideband, $f_m$ is the corresponding sideband frequency, $f_c$ is the frequency of the central peak, $B_2$ is the total resonator dispersion and $\tau$ is the pulse width (for a pulse given by $\text{sech}^2\qty(\frac{t}{\tau})$, the FWHM is $1.763\tau$). By fitting to $(f_m-f_c)^2$ as a function of $m$ we obtain a total resonator dispersion of $\qty(-0.44 \pm 0.02)\times10^{-24}\,\text{s}^2$. If we assume the total resonator dispersion is due entirely to the fiber, dividing this result by the 2.3\,m fiber length gives the group velocity dispersion $\beta_2$ of the fiber as $\qty(-0.191 \pm 0.009)\times10^{-24}\,\text{s}^2\,\text{m}^{-1}$. This agrees with the value of $-0.181\times10^{-24}\,\text{s}^2\,\text{m}^{-1}$ we calculate from numerical fiber eigenmode analysis \cite{marcuseInterdependenceWaveguideMaterial1979}. The refractive indices used in this calculation were obtained from data provided by fiber manufacturer.

The pulse energy and duration of solition lasers are often estimated using the soliton area theorem \cite{hausModelockingLasers2000}:
\begin{equation}
    E\tau = \frac{2\abs{\beta_2}}{\gamma},
\end{equation}
where $\beta_2$ is the group velocity dispersion, $\gamma$ is the nonlinear coefficient and $\tau$ is the pulse duration ($\text{FWHM}=1.763\tau$). We calculate $\gamma =$ $0.83\times10^{-4}\,\text{W}^{-1}\text{m}^{-1}$ using a mode field radius of 12.0\,\textmu m and a nonlinear refractive index of $2.1\times10^{-20}\,\text{m}^2\text{W}^{-1}$ \cite{aggerSupercontinuumGenerationZBLAN2012}. We use the value of $-0.191\times10^{-24}\,\text{s}^2\text{m}^{-1}$ for $\beta_2$ that was calculated earlier from the Kelly sidebands present in the optical spectrum. Using the measured pulse energy of 3.2\,nJ and accounting for the 77\% reflectivity output coupler mirror, the soliton area theorem gives a pulse duration of 590\,fs. Although this agrees well with our measured value of 580\,fs, this is still only an approximation as these are not strictly solitons but are more accurately described as average solitons \cite{kellyAverageSolitonDynamics1991}. In this operating regime, the pulse characteristics are determined by the balance between the accumulated phase in a resonator round trip due to the averaged dispersion and nonlinearity \cite{woodwardDispersionEngineeringModelocked2018}. The pulse energy varies significantly as it propogates around the resonator due to gain/loss and the average is markedly lower than that at the output end of the fiber. The pulse duration and spectral width however, remain relatively constant throughout a resonator round trip, based on numerical simulations by Woodward \cite{woodwardDispersionEngineeringModelocked2018}. The pulses therefore disagree with the soliton area theorem at many points in the resonator such that the averaged effects of self phase modulation balance anomalous dispersion.

Experiments using 72\% and 87\% reflectivity output coupler mirrors resulted in mode-locked pulses with different durations and energies. The results are summarised in Table \ref{tab:results summary}. For the 87\% reflectivity mirror, instabilities in the mode-locked output limited the incident pump powers. The optical spectrum in this case contained stronger sidebands, up to ten times the intensity of the central operating wavelength. The larger output coupling fraction of the 72\% reflectivity mirror resulted in the greatest pulse enrgy of 3.6\,nJ.

\begin{table}[htbp]
\centering
\caption{Measured parameters for mode-locked operation with various output coupler mirrors.}
\label{tab:results summary}
\begin{tabular}{ccccccc}
\hline
R {[}\%{]} & $\text{P}_1$ {[}W{]} & $\text{P}_2$ {[}W{]} & Power {[}mW{]} & Energy {[}nJ{]} & Duration {[}fs{]} & Peak power {[}kW{]} \\ \hline
72 & 8.0 & 6.1 & 244 & 3.6 & 650 & 5.5 \\
77 & 6.8 & 5.5 & 216 & 3.2 & 580 & 5.5 \\
87 & 6.2 & 5.3 & 141 & 2.1 & 600 & 3.5 \\ \hline
\end{tabular}
\end{table}




Our NPR mode-locking results presented in this work significantly improve upon previous mode-locking demonstrations at this wavelength because the shorter, sub-picosecond pulses result in a greater peak power. We have shown that NPR is a robust method for short pulse generation well into the mid-infrared. Our peak power and pulse duration are comparable to those demonstrated in similar experiments at 2.8\,\textmu m and 3.1\,\textmu m, respectively, even though the erbium 3.5\,\textmu m transition is a lower gain lasing transition.

Useful information is also obtained about the fluorozirconate fiber. The measured group velocity dispersion begins to create a useful set of parameter values for this wavelength range when combined with the measured values from similar experiments. These are summarised in Table \ref{tab:beta summary}.
\begin{table}[htbp]
\centering
\caption{Measured values of the group velocity dispersion from NPR mode-locking experiments in the mid-infrared.}
\label{tab:beta summary}
\begin{tabular}{ccc}
\hline
Wavelength {[}\textmu m{]} & $\beta_2$ {[}$\text{s}^2\text{m}^{-1}${]} & Ref. \\ \hline
2.8 & 0.094 & \cite{duvalFemtosecondFiberLasers2015}    \\
3.1 & 0.120 & \cite{wangUltrafastDy3Fluoride2019}    \\
3.5 & 0.191 & This work    \\ \hline
\end{tabular}
\end{table}

The erbium 3.5\,\textmu m transition has previously been shown to have a tunability of over 450\,nm in CW operation from 3.33\,\textmu m and almost reaching 3.8\,\textmu m \cite{henderson-sapirVersatileWidelyTunable2016}. Future experiments with increased pump powers and dispersion management should therefore be able to achieve further spectral broadening and hence shorter pulses while still allowing for wavelength tuning, making this system more appealing for spectroscopy applications.

We have demonstrated mode-locked operation of a fiber laser at 3.5\,\textmu m using NPR. The average output power and repetition rate were 216\,mW and 68\,MHz, corresponding to a pulse energy of 3.2\,nJ. Autocorrelation measurements gave a pulse duration of 580\,fs, the first sub-picosecond pulses from a fiber laser operating directly at this wavelength. The resulting peak power for the laser was 5.5\,kW.

\section*{Acknowledgments}
The authors thank Matthew R.\ Majewski and Robert I.\ Woodward for helpful discussions.
The authors also acknowledge the expertise, equipment, and support provided by the Australian National Fabrication Facility (ANFF) at The University of Adelaide.
This research was supported by the Australian Research Council, the Asian Office of Aerospace R\&D.

\section*{Funding}
Asian Office of Aerospace R\&D (AOARD) Grants FA-9550-20-1-0160 and FA2386-19-0043.

\section*{Diclosures} 
OHS: Mirage Photonics (I,P)
 

\bibliography{Bawden.bib}

\end{document}